\documentclass[11pt]{article}
\usepackage{amsmath,amssymb,cite}
\usepackage{hyperref}

\numberwithin{equation}{section}

\begin{document}

\begin{center}
\large{\bf Quantum Physics and Fluctuating Topologies: Survey}

\bigskip
         \begin{small}
         {\bf M. Asorey$^{a}$\footnote{asorey@unizar.es}, 
         A. P. Balachandran$^{b,c}$\footnote{balachandran38@gmail.com},\, 
         G. Marmo$^d$\footnote{marmo@@na.infn.it},\, 
         I. P. Costa e Silva$^e$\footnote{ivanpcs@mtm.ufsc.br}, \\
         A. R. de Queiroz$^{f}$\footnote{amilcarq@unb.br}\,
	 P. Teotonio-Sobrinho$^{g}$\footnote{teotonio@if.usp.br} \,and\,
 	 S. Vaidya$^h$\footnote{vaidya@cts.iisc.ernet.in}} \\
\vspace{.25cm}
$^a${\it Departamento de F\'{\i}sica Te\'orica, Facultad de Ciencias,
Universidad de Zaragoza, 50009 Zaragoza, Spain} \\
\smallskip
$^b${\it Department of Physics, Syracuse University, Syracuse, N. Y. 13244-1130, USA}\\
\smallskip
$^c${\it Institute of Mathematical Sciences, Chennai, India}\\
\smallskip
$^d${\it Dipartimento di Scienze Fisiche, Universit\`a di Napoli Federico II and INFN Sezione di Napoli, 
Via Cintia, 80126, Napoli, Italy}\\
\smallskip
$^e${\it Departamento de Matem\'atica, Universidade Federal de Santa Catarina, 88040-900, 
Florian\'opolis, SC, Brazil}\\
\smallskip
$^f${\it Instituto de Fisica, Universidade de Brasilia, Caixa Postal 04455, 70919-970, Brasilia, DF, Brazil}\\
\smallskip
$^g${\it Instituto de F\'{\i}sica, Universidade de S\~ao Paulo, Caixa Postal 66318, 05315-970, S\~ao Paulo, SP, Brazil}\\
\smallskip
$^h${\it Centre for High Energy Physics, Indian Institute of Science, Bangalore, 560012, India}
\end{small}
\end{center}

\begin{abstract}
The spin-statistics connection, quantum gravity and other physical considerations suggest that classical 
space-time topology is not an immutable attribute and can change in quantum physics. The implementation 
of topology  change  using quantum principles has been studied for over two decades by a few of us. There 
has been a recent revival of interest in some of our work, dating back to as early as 1995. The present paper 
is meant as a resource article to our major relevant papers. It contains summaries of the contents of the 
cited papers and the corresponding links wherever available.
\end{abstract}

\section{Introduction}

The role of classical topology in quantum physics is not well-understood even after nearly one hundred 
years since the formulation of the principles of quantum theory. Observables in quantum physics are 
self-adjoint operators and their spectra, and the mode of emergence of space-time from such data 
is a central problem of  quantum epistemology.  A few of us have been collaborating on this topic since 
late 1970's  and have in fact coauthored two books on this subject 
\cite{balachandran1983gauge,balachandran1991classical}.

A related issue concerns quantum gravity. It is widely speculated that space-time topology can change in 
quantum gravity  and classical manifolds are just approximate idealizations emergent from  quantum states
 which are localized in neither one nor another topological space. But this idea has remained speculative. 
The tools  tried for its implementation in quantum gravity have been functional integrals. But the latter are 
ill-defined on smooth Lorentzian manifolds \cite{Geroch1967} and anyway the Lorentz metric 
practically excludes topology change because of known theorems on cobordism of manifolds 
\cite{Geroch1967,Sorkin1986-IJMP,Dowker:1997hj,Borde:1999md,Dowker:1999wu,GSW}. 
The use of Euclidean functional integrals for gravity are fraught with problems and cannot be regarded 
as reliable  probes of  so fundamental an issue.

It is in this context that a few of us began the  study of simple quantum mechanical systems and formulate 
topology change using properties of wave functions.This was in 1995, some 17 years ago 
\cite{Balachandran1995c}. Recent work \cite{Shapere2012b} shows a revival of interest in this paper and 
also in earlier studies of classical topology in quantum physics by our group of collaborators. With this in 
mind, we have collected together here some of our relevant publications. They are not exhaustive as 
regards our work, but do contain adequate citations to follow the trails of our thought.

We have divided this collection into two parts. It is the first which works with systems with finite degrees 
of freedom and uses basic ideas on differential operators such as the Laplacian on manifolds with boundaries 
to explore spatial topology and and its transmutation. The second is also involved with spatial topology. In 
one approach reported here, an algebraic approach extracted from large distance physics is developed for 
this purpose. It also contains a paper on quantum fields on curved spaces with nontrivial topology and shows 
how diffeomorphisms can change the domain of the field Hamiltonian and become anomalous.

We now give brief introductions to the papers.

\section*{Part 1. Topology Change and Quantum Physics}

\subsection{Topology Change and Quantum Physics}

\begin{enumerate}
  \item A. P. Balachandran,  G. Bimonte, G. Marmo and A. Simoni, \emph{Topology change and quantum physics} {\bf Nucl. Phys.}, B446 (1995) \cite{Balachandran1995c}

In this paper, the role of topology in elementary quantum physics is explored. The attributes of classical 
spatial topology are shown to emerge from properties of state vectors with suitably smooth time evolution. 
Equivalently, they emerge from the domain of the quantum Hamiltonian, which is often specified by boundary 
conditions . Several examples are presented where classical topology is changed by smoothly altering the 
boundary conditions. When the parameters labeling the latter are treated as quantum variables, quantum 
states need not give a well-defined classical topology, instead they can give a quantum superposition of 
such topologies.  Arguments of Sorkin based on the spin-statistics connection which indicate the need for 
topology change in quantum gravity and further arguments presented here suggest that Einstein gravity and 
its minor variants are effective theories of a deeper description with additional novel degrees of freedom.

%\end{enumerate}

\subsection{Global Theory of Quantum Boundary Conditions and Topology Change}

%\begin{enumerate}

\item M. Asorey, A. Ibort and G. Marmo, \emph{Global Theory of Quantum Boundary Conditions and 
Topology Change},  {\bf Int.J.Mod.Phys. A 20}, 1001 (2005)  \cite{Asorey2005}. 
  
\item M. Asorey, A. Ibort and G. Marmo, \emph{The Boundary Grassmannian for Dirac Operators} 
(in preparation).
  
\item A. Ibort, \emph{Three lectures on global boundary conditions and the theory of self--adjoint extensions 
of the covariant Laplace--Beltrami and Dirac operators on Riemannian manifolds with boundary}, 
{\bf AIP  Conf. Proc.}, 1460 (2012) \cite{Ibort2012a}.

\item M Asorey, D Garcõa-Alvarez and J M Munoz-Castaneda,
\emph{Vacuum energy and renormalization on the edge}, {\bf J. Phys. A 40}, 6767 (2007)
\cite{Asorey:2007rt}

These four are companion papers to the preceding one. In the first, the theory of the self-adjoint Laplacians 
on a Riemannian manifold with 'regular' boundary is systematically treated. The possible boundary conditions 
are fully classified and their equivalence with von Neumann's conditions are established. This analysis and 
the corresponding analysis of Dirac operators in the second paper are particularly adapted to discuss topology 
change. A striking result, with implications for topology change, is that for small deformations of the Dirichlet 
boundary condition, the Laplacian has infinitely many negative energy states localized near the boundary.

The second paper is yet to be published, but its contents as well as those of the first paper are summarized 
in the third. The fourth paper is particularly important for its treatment of the dynamics of topology change 
and the boundary renormalization group flow.

\subsection{Quantum Boundary Conditions and Spectral Action Principle}

\item L. C. de Albuquerque, P. Teotonio-Sobrinho and S. Vaidya, \emph{Quantum topology change and large 
N gauge theories}, {\bf JHEP}, 0410 (2004) \cite{Albuquerque2004}.
  
This paper addresses the question of recovering usual topologies from dynamics on the space of 
boundary conditions $Q_{TOP}$. The explicit model at hand is a one-dimensional model consisting of 
a collection of $N$ unit segments, generalizing the original example of Balachandran et al 
\cite{Balachandran1995c}. Among all possible boundary conditions, there are special ones that glue the 
line segments into circles of various sizes. These points of $Q_{TOP}$ correspond to a classical topology. 
The formalism of spectral triples in noncommutative geometry then provides the motivation for dynamics 
on $Q_{TOP}$, thus making the system closely analogous to Euclidean quantum gravity for fluctuating 
topologies.

The spectral action governing the dynamics on $Q_{TOP}$ corresponds to a $U(N)$ lattice gauge theory with 
a single plaquette where the holonomy $g$ characterizes the boundary condition. The partition function has 
a single free parameter $\beta$ as usual. We show that topology becomes classical and localized in the limit 
$\beta \rightarrow \infty$ for all values of $N$. For large $N$ the system has a third order phase transition at 
$\beta_c=1$. Although topology is not exactly localized for finite $\beta$, one of the phases imposes 
constraints on classical topology. It turns out that for $\beta > \beta_c$, the only classical circles that can 
occur have finite sizes. In particular, the probability distribution for large $\beta$ is dominated by classical 
circles of unit length.

\end{enumerate}
  
\section*{Part 2: Quantum Geons: Spin-Statistics, Topology Change, Anomalies} 

\begin{enumerate}
  
\item A. Balachandran, E. Batista, I. Costa e Silva and P. Teotonio-Sobrinho, \emph{Quantum topology 
change in (2+1)-dimensions}, {\bf Int. J. Mod. Phys.}, 2000, A15, 1629 (2000) \cite{Balachandran2000}.
  
\item A. P. Balachandran, E. Batista, I. Costa e Silva and P. Teotonio-Sobrinho, \emph{The 
Spin-Statistics Connection in Quantum Gravity}, {\bf Nucl. Phys.}, B566, 441 (2000) \cite{Balachandran2000a}.

\item A. P. Balachandran, E. Batista, I. Costa e Silva and P. Teotonio-Sobrinho,  \emph{A Novel 
Spin-Statistics Theorem in (2 + 1)d Chern-Simons Gravity}, {\bf Mod. Phys. Lett.}, A16, 1335 (2001) 
\cite{Balachandran2001}.
   
In these  papers, instead of adopting dynamical boundary conditions for standard operators like the Laplacian,
an algebra of operators capturing topological degrees of freedom is introduced as follows. Starting from 
a Yang-Mills-Higgs theory with gauge group $G$ on an oriented, non-compact Riemannian surface $\Sigma$ 
of genus $g$ and one asymptotic region, one assumes further that the symmetry is broken down to a 
finite subgroup $H \subset G$, and considers the low energy limit of the theory.The configuration space 
then splits into disjoint sectors.They are characterized by  vortices with fluxes valued in H. Introducing also 
large diffeomorphisms which can change these fluxes, one gets a topological algebra suitable for large 
distance physics. It is quantized, giving a description of single vortex irreducible representations (IRR's). 
Each such IRR describes a single irreducible vortex. Its representation on the  two -vortex vector states is 
then given by a 'coproduct' (the algebras being Hopf \cite{deWildPropitius:1995cf}). This representation can 
be decomposed in turn into IRR's and in general contains many IRR's even when the starting single 
vortex transforms irreducibly. Each such IRR is interpreted as describing a composite irreducible vortex. 
In this way, we have an effective description of  two (or many) irreducible vortices fusing and changing 
into single irreducible vortices, and that is topology change.
 
Each genus $g$ surface supports a solitonic excitation called a geon, whose quantum attributes were 
first studied by Friedman and Sorkin \cite{Friedman:1980st,Friedman:1982du}. In the absence of 
topology change, geons can violate the usual spin-statistics connection. In later work by us, the above ideas 
on topology change were applied to these geon manifolds as well, thereby achieving topology change. A 
new spin-statistics theorem for these two-dimensional geons, which differs from the standard one, is 
also proved.

\item A. P. Balachandran and A. R. de Queiroz. \emph{Quantum Gravity: Mixed States from 
Diffeomorphism Anomalies}, {\bf JHEP}, 11 (2011) \cite{Balachandran:2011gj}.

As alluded to above, geons are solitonic excitations on spatial manifolds (of any dimension) with 
non-trivial topology with remarkable properties. They were first investigated in the context of quantum gravity 
by Friedman and Sorkin. 

Now in a manifold with geons, a `large' diffeomorphism (an element of the mapping class group) may 
change the domain of a self-adjoint operator such as the Laplacian. If it does so, it is anomalous just as 
axial `symmetry' transformations in QCD are anomalous. This paper studies this question by examining 
the action of  the mapping class group on the fundamental group of  manifolds with geons. In this manner, it 
is established that the  mapping class group can in fact become anomalous. When it does so, conventional 
quantum gravity based on pure states loses diffeomorphism invariance. The paper also discusses how the 
use of suitable mixed states \cite{Balachandran:2011bv} may enable the construction of an anomaly-free 
quantum gravity.
  
\end{enumerate}

%\section{Acknowledgements}
%APB is supported by the Institute of Mathematical Sciences, Chennai. ARQ is supported by CNPq under 
%process number 307760/2009-0.

\bibliographystyle{JHEP}

\providecommand{\href}[2]{#2}\begingroup\raggedright

\end{document}